\documentclass[aps,preprint,floats]{revtex4}
\usepackage{epsfig}
\usepackage{epsf}
\usepackage{color}
\usepackage{hyperref}

\newcommand{\rr}{{\bf r}}
\newcommand{\rrp}{{\bf r}^\prime}

\newcommand{\oogut}{{\"{O}\u{g}\"{u}t }}

\begin{document}

\title{First-principles GW-BSE excitations in organic molecules}

\author{Murilo L. Tiago$^{(a)}$\footnote{corresponding author, Tel.:
    +1 512 232 2085, fax: +1 512 471 8694} and James R. Chelikowsky$^{(a,b)}$} 
\affiliation{
  $^{(a)}$ Department of Chemical Engineering and Materials Science,
  Institute for the Theory of Advanced Materials in Information
  Technology, Digital Technology Center, University of Minnesota,
  Minneapolis, MN 55455, USA.\\
 $^{(b)}$ Departments of Physics and Chemical Engineering, Institute
  for Computational Engineering and Science, University of Texas,
  Austin, TX 78712, USA.}

\date{\today}

\begin{abstract}
We present a first-principles method for the calculation of optical 
excitations in nanosystems. The method is based on solving the Bethe-Salpeter
equation (BSE) for neutral excitations. The electron self-energy is evaluated
within the GW approximation, with dynamical screening effects described
within time-dependent density-functional theory in the adiabatic, local
approximation. This method is applied to two systems: the benzene
molecule, C$_6$H$_6$, and azobenzene, C$_{12}$H$_{10}$N$_2$. We
give a description of the photoisomerization process of azobenzene after an
$n-\pi^\star$ excitation, which is consistent with multi-configuration 
calculations.
\end{abstract}

\maketitle

\newpage

\section{Introduction}

Predicting many-body excitations of weakly correlated systems from first
principles is important, both to help understand the behavior of known 
systems as well as to design novel materials and technological
devices. The study of optical excitations in systems at nanoscale size
has been accentuated by recent achievements in nanotechnology and the
potential growth of the field \cite{ichimura00,hugel02,norikane04}. 
On the theory side, time-dependent density functional theory in the
adiabatic, local approximation (TDLDA) has been 
used to study a wide class of nanosystems ranging from organic
molecules to semiconducting and metallic clusters (see {\it e.g.}
\cite{vasiliev02} and references therein). Alternatively, optical
excitations of atoms, small molecules and bulk materials have been
studied from first principles by solving the Bethe-Salpeter equation
(BSE) for electrons and holes \cite{rohlfing00}. In bulk materials,
the BSE approach is known to fully describe excitonic effects, which
are missing in TDLDA \cite{rohlfing00,onida03}. In confined systems 
such as clusters and isolated molecules, the BSE approach is
expected to be more accurate than TDLDA, but extensive comparisons of
both methods have not been done so far due to the complexity of most
numerical implementations of the BSE. One source of complexity is the
explicit evaluation of the dielectric function of the system. Early
implementations \cite{rohlfing00} use a Fourier expansion of the
static dielectric function, with dynamical effects included via
semi-empirical models or ignored altogether. More recent studies
employ generalized plasmon-pole models in the description of dynamical
screening \cite{benedict03}. \oogut and collaborators
\cite{ogut03} have presented a fully {\it ab initio} technique for the
calculation of the static dielectric function in real space, leading to 
important simplifications compared to Fourier-expansion techniques.

We propose a formulation of the BSE method that avoids
the explicit evaluation of the dielectric function. All the
information about electronic screening is contained in the
polarizability operator, which is calculated within TDLDA by solving a
generalized eigenvalue problem \cite{vasiliev02,casida95}. After
numerical diagonalization, eigenvectors give the spatial dependence of
the polarizability, while eigenvalues give its frequency (time)
dependence. Both the electronic self-energy and electron-hole interaction
kernel, which enter explicitly in the BSE, are computed directly from
the TDLDA polarizability. Once the BSE is reduced to a generalized,
frequency-dependent eigenvalue problem, it is numerically diagonalized. The
resulting normal modes are associated to neutral, optical excitations
of the electronic system.


\section{Theoretical Method}

The underlying description of the electronic system is obtained within
density-functional theory (DFT). The Kohn-Sham equations are solved in 
real space. Degrees of freedom related to core electrons are excluded 
from the problem by using norm-conserving
pseudo-potentials. We start by expressing the random-phase
approximation (RPA) polarizability operator in the space of Kohn-Sham
eigenstates in the usual way:

\begin{equation}
P_{cv,c'v'}^0 (\omega) =  {
\delta_{c,c'} \delta_{v,v'}  \over
\omega - \omega_{cv} + i0^+ } - {
\delta_{c,c'} \delta_{v,v'}  \over \omega +
\omega_{cv} - i0^+ }
\label{e_rpa}
\end{equation}
where $\omega_{cv}$ is the difference between two Kohn-Sham
eigenvalues, $\varepsilon_c-\varepsilon_v$, and $0^+$ denotes an
arbitrarily small quantity. We assume $\hbar = 1$ and ignore spin
indices for the sake of clarity. The case of spin-polarized systems
require slight modifications in the formalism. Here, we denote
occupied Kohn-Sham orbitals by $v,v'$ and unoccupied (virtual) ones by
$c,c'$. Generic orbitals are denoted by letters $i,j,n$. Appropriate
occupation factors should be included in Eq. (\ref{e_rpa}) if
partially populated orbitals are present \cite{casida95}. Within
TDLDA, the polarizability is written in terms of eigenvalues
$\{\omega_m\}$ and eigenvectors $\{F_{vc}^m\}$ of an effective
eigenvalue problem \cite{casida95},

\begin{equation}
\Pi_{cv,c'v'} = \sum_{vc,v'c',m} \left[ {{F_{vc}^m}^\star
F_{v'c'}^m \over \omega - \omega_m + i0^+ } - {F_{vc}^m
{F_{v'c'}^m}^\star \over \omega + \omega_m - i0^+ } \right] \;\;.
\label{e_tdlda}
\end{equation}
Both polarizabilities can also be expressed in real space by including
Kohn-Sham eigenvectors according to the rule

\begin{equation}
\left\{ {\cal O} \right\} (\rr,\rrp) = \sum_{ij,i'j'} u_i(\rr)
u_j^\star(\rrp) \left\{ {\cal O} \right\}_{ij,i'j'}
u_{i'}^\star(\rr) u_{j'}(\rrp) \;\; .
\label{e_representation}
\end{equation}

One key quantity in solving the Bethe-Salpeter equation for optical
excitations is the electronic self-energy $\Sigma$, which can be computed from
first principles within the so-called GW approximation (GWA)
\cite{hybertsen86,aulbur00}. This method has been used in a
number of different systems and at different levels of
sophistication. At its lowest level, the self-energy is schematically
evaluated as $\Sigma_0 = G_0 W_0$, where $G_0$ is the (DFT) Green's
function and $W_0 = [1 - VP_0]^{-1} V$ is the screened Coulomb
interaction, with $V$ being the bare Coulomb interaction, $V(\rr) =
e^2/r$. In this implementation, we go beyond RPA and use the TDLDA
polarizability, Eq. (\ref{e_tdlda}). Having in mind that $\Pi$ and $P_0$
are related to each other via $\Pi^{-1} = P_0^{-1} - (V+f)$
\cite{onida03,hanke78}, where $f$
is the TDLDA interaction kernel, $f = {\delta V_{xc} \over \delta n}$,
one can arrive at this expression for the self-energy:

\begin{widetext}
\begin{equation}
\Sigma (\omega^\prime) = i \int { {\rm d} \omega \over 2 \pi}
e^{-i\omega 0^+} G_0 (\omega^\prime-\omega) 
\left[ V + V \Pi (\omega) V + {1
\over 2} V \Pi (\omega) f + {1 \over 2} f \Pi(\omega) V \right] \;\;.
\label{e_sigma}
\end{equation}
\end{widetext}

In Eq. (\ref{e_sigma}), the first term inside square brackets denotes
the bare exchange (Hartree) contribution. The second term is the
correlation part. The last two terms have vertex corrections. 
Some aspects of Eq. (\ref{e_sigma}) should be pointed out:

\begin{enumerate}
\item The additional vertex terms are written in symmetrized form, so
  that the resulting self-energy is symmetric with the interchange of
  arguments. In many-body notation, $\Sigma(1,2) = \Sigma(2,1)$.
\item By ignoring the $f$ kernel and evaluating the polarizability as
  $\Pi^{-1} = P_0^{-1} - V$, we recover the self-energy at $\Sigma_0$
  level of approximation. Adding a $f$ kernel in the polarizability amounts
  to enhanced screening, which is partially compensated by the
  inclusion of vertex terms.
\item The frequency dependence of
the polarizability is known from Eq. (\ref{e_tdlda}). Plasmon pole
models are not needed. 
\end{enumerate}

Numerically, matrix elements of the self-energy
can be evaluated by expressing the quantities $V,\Pi,f$ in
representation of Kohn-Sham orbitals using Eq. (\ref{e_representation})
and evaluating the self-energy directly from Eq. (\ref{e_sigma}). We do not
impose self-consistency between polarizability, self-energy and
Green's function.

Optical excitations can be obtained either from the singularities of
the TDLDA polarizability in frequency domain, Eq. (\ref{e_tdlda}), or
by solving the BSE. In the second case, we set up a generalized
eigenvalue problem where the interaction kernel $K$ is no longer of
the form $K = V + f$ as in TDLDA, but given by $K = {\delta (V_H +
\Sigma) \over \delta G}$ \cite{rohlfing00}. From Eq. (\ref{e_sigma}),
this reduces to

\begin{equation}
K_{vc,v'c'} = 
\left[ V \right]_{vc,v'c'} +  \left[ V + V \Pi
V + {V \Pi f  \over 2} + { f \Pi V \over 2}
\right]_{vv',cc'} \;\; .
\label{e_kernel}
\end{equation}
With this kernel operator, the BSE is solved numerically using standard
procedures. The term inside the first pair of square brackets should
be ignored for spin-triplet excitations \cite{rohlfing00,onida03}.

Having a frequency-dependent kernel makes the BSE a non-standard
eigenvalue equation. In periodic systems, dynamical effects were found
to be negligible if one is interested in the linear optical spectrum.
In finite systems, quantum confinement effects increase the strength
of the interaction kernel, resulting in somewhat stronger but still
limited dynamical effects (about 0.01 eV for the SiH$_4$
system)\cite{rohlfing00}. Although we ignore those
effects in the present work, they can be included by 
evaluating polarizability and kernel at some
frequency close to the actual excitation frequencies obtained by
solving the BSE.


\section{Applications}

A test case of the procedure above is the isolated benzene
(C$_6$H$_6$) molecule. For this system, we have solved the Kohn-Sham
equations on a regular grid, with grid spacing 0.4 a.u. (0.21
$\AA$). Electronic wave-functions were required to vanish outside a
sphere of radius 16 a.u. (8.47 $\AA$), centered on the
molecule. Carbon-carbon and carbon-hydrogen bond lengths were fixed at
their experimental values. After including
self-energy corrections, the ionization potential is found to be 9.30
eV, in good agreement with the experimental value of 9.3 eV
\cite{herzberg}. Benzene has a negative electron affinity of -1.12 eV,
measured in resonance scattering experiments \cite{burrow86}, 
which is in excellent agreement with the calculated electron affinity 
of -1.11 eV. 
Here, the negative electron affinity is interpreted as energy of the
$E_{2u}$ resonant state of the anion, and not the lowest energy
required to extract one electron from the anion (which is zero because
the anion is unstable). We observed that the inclusion of vertex
corrections together
with a TDLDA polarizability is essential for an accurate prediction of
the ionization potential and electron affinity of benzene. In fact,
the $\Sigma_0$ approximation predicts those quantities to be 9.84 eV
and -0.54 eV respectively. Defining a ``HOMO-LUMO gap'' as the
difference between ionization potential and electron affinity, the
present approach and the $\Sigma_0$ approach agree in the value of the
gap: 10.4 eV. This is somewhat consistent with the observation by del
Sole and collaborators \cite{delsole94}, who conducted a similar
analysis in bulk silicon and found significant shifts in the
valence band maximum and conduction band minimum, but only small change
in the energy gap.

The low-energy optical spectrum of this molecule is 
dominated by a $\pi-\pi^\star$ complex, which arises from transitions 
between the degenerate highest occupied molecular
orbital (HOMO) and lowest unoccupied molecular orbital (LUMO). Table
\ref{t_benzene} shows a comparison between TDLDA and BSE predictions
for some excitations in the $\pi-\pi^\star$-complex. Although singlet
transitions $B_{1u}^1$ and $E_{1u}^1$, which are the dominant ones in the
low-energy part of the absorption spectrum, are equally well described 
by both methods, there
is a small but well-defined blue shift of dark transitions $B_{1u}^3$
and $B_{2u}^1$ within TDLDA. With the exception of the $E_{1u}^1$
transition, excitation energies obtained by solving
the BSE are typically underestimated with respected to measured
quantities. The overall deviation between measured and calculated
transition energies is 0.1 to 0.3 eV.

The absorption spectrum of benzene, Fig. \ref{f_benzene}, shows
the bright $E_{1u}^1$ transition, at 7.0 eV. The low,
flat feature in the 6.0-7.0 eV range is due to coupling between
$\pi-\pi^\star$ transitions and
vibrational modes \cite{herzberg,bertsch01}, and it is absent in the
calculated spectra because of the assumed structural rigidity. Beyond
10 eV, a number of sharp features in the measured spectrum results from
transitions involving Rydberg states \cite{herzberg,koch72}. The
limited numerical accuracy in that energy range prevents a detailed
identification of such transitions in the calculated spectrum.



A more interesting system is the azobenzene (C$_{12}$H$_{10}$N$_2$)
molecule. This molecule has attracted considerable attention recently
due to exciting experiments of light-driven mechanical manipulation at
nanoscale. Some of these experiments are: demonstration of
light-driven manipulation of liquid droplets on an optically active
surface \cite{ichimura00}; energy conversion in a polymeric azobenzene
chain \cite{hugel02}; synthesis of an azobenzene compound which
exhibits hinge-like motion when photo-irradiated \cite{norikane04}.
The basic phenomenon explored in these experiments
is photo-isomerization: the structure of azobenzene is induced to change
from a {\it trans-}azobenzene configuration (TAB) to a metastable {\it
cis-}azobenzene configuration (CAB) upon absorption/emission of
radiation with the appropriate frequency. On the theory side, the
detailed structural distortion involved in the process has been subject of
discussion. Early analyses based on a restricted configuration
interaction (CI) calculation \cite{monti82} indicated that photo-isomerization
following excitation to the first excited state occurs via
inversion of one nitrogen in the azo group, causing an increase of
$120^o$ in the C-C-N bond angle (``inversion path''). On the other
hand, recent multi-configuration and TDLDA calculations
\cite{cattaneo99,tiago} indicate that
the most favorable path is a torsion of the N-N bond around its axis,
causing a $180^o$ rotation of one phenyl group with respect to the
other (``rotation path''). Diau has proposed an alternative
``synchronous inversion path'' that can also lead to
photo-isomerization \cite{diau04}.

We have studied optical excitations in the isolated azobenzene
molecule, in order to test the accuracy of the present BSE
approach. Isomerization along the rotation path was simulated by initially
fixing the CNNC dihedral angle and minimizing the DFT total
energy of the system with respect to all other degrees of freedom. A
sequence of configurations was then obtained for various choices of
the dihedral angle ranging from 0$^o$ (CAB) to 180$^o$ (TAB), as discussed
in \cite{tiago}. The Kohn-Sham equations were solved on real space,
using a boundary radius of 20 a.u. (10.58 $\AA$) and grid spacing 0.4
a.u. (0.21 $\AA$). At the
equilibrium configuration, {\it trans-}azobenzene has ionization
potential 8.5 eV \cite{natalis81}, which is consistent
with the corresponding values of 8.6 eV, obtained from GW with
a self-energy given by Eq. (\ref{e_sigma}).

The calculated potential energy of the system in its ground state and
the two lowest excited states is shown in
Fig. \ref{f_azo_singlet}. At each configuration, the potential energy is
computed as a sum of ground-state potential energy and excitation
energy, the latter computed either within TDLDA or BSE. Calculated
excitation energies for the CAB and TAB configurations are shown in
Table \ref{t_azo}. Deviations between the BSE and experimental
excitation energies are small, within 0.3 eV. The CAB $\pi\pi^\star$
excitation energy 
has larger deviation, of almost 0.8 eV, which may indicate a
difficulty in the theoretical description of higher-energy
excitations. On the other hand, excitation energies obtained within
TDLDA are systematically lower than the ones obtained within BSE by more
than 0.4 eV, and substantially different from measured excitation energies.

Figure \ref{f_azo_singlet} shows that, once the system absorbs radiation
and is promoted to its first excited state at either CAB or TAB
configurations, there is no energy barrier preventing fast relaxation
to an intermediate configuration, with dihedral angle around
$90^o$. From that configuration, the system can decay to ground state
and relax back to either the CAB or TAB configurations. This mechanism
allows $trans \longrightarrow cis$ or $cis \longrightarrow trans$
isomerization after excitation to state $n\pi^\star$ (the first excited 
state). Excitation to the second excited state is not
found to induce isomerization, unless there is decay to state $n\pi^\star$,
since the profile of potential energy is rather flat
along the rotation path.

Photo-isomerization involving higher-order excited states, which has
been observed in laboratory, can also be studied once four-particle
and more complex excitations are included in the
theory. Multi-configuration calculations
\cite{cattaneo99,fujino01} indicate that an excited state
composed of two electrons promoted from the HOMO to the LUMO has a
barrier-less potential energy profile along the rotation path. That
excited state is not accurately described within either TDLDA or BSE
approaches in the current two-particle approximation. Despite this
limitation, excited states of azobenzene can be studied within the BSE
approach or, to a lower level of accuracy, TDLDA.


\section{Conclusion}

In summary, we have discussed a first-principles method for the calculation
of optical excitations in nanosystems, with applications to two organic
molecules: benzene and azobenzene. The method is based on solving the 
Bethe-Salpeter equation (BSE) for neutral excitations, with dynamical
screening described within TDLDA. To our knowledge, this is the {\it first
time} that optical excitations in organic molecules are calculated using
Green's function-based methods. The approach is completely system-independent
and it can also be used to study inorganic systems.
This work was supported in part by the National Science Foundation under
DMR-0130395 and DMR-0325218 and the U.S. Department of Energy under
DE-FG02-89ER45391 and DE-FG02-03ER15491. The calculations were
performed at the Minnesota Supercomputing Institute and at the
National Energy Research Scientific Computing Center (NERSC).



\newpage

\begin{table}[t]
\caption{Excitation energies of the lowest-energy neutral excitations
  in benzene. The BSE excitation energies were obtained with a static
kernel. Energies in eV.}
\label{t_benzene}
\begin{tabular}{lcccc} \hline \hline
 & & \hspace{0.3cm} TDLDA \hspace{0.3cm}
 & \hspace{0.3cm} BSE \hspace{0.3cm}
 & \hspace{0.3cm} Exp. \cite{doering69} \hspace{0.3cm} \\ \hline
triplet & $B_{1u}^3$ & 4.5 & 3.6 & 3.9 \\ \hline
singlet & $B_{2u}^1$ & 5.4 & 4.9 & 5.0 \\
\hspace{0.5cm} & $B_{1u}^1$ & 6.2 & 6.1 & 6.2 \\
 & $E_{1u}^1$ & 6.9-7.2 & 7.2 & 6.9 \\\hline \hline
\end{tabular}
\end{table}

\begin{table}[t]
\caption{Excitation energies of the lowest-energy neutral excitations
  in azobenzene. Energies in eV.}
\label{t_azo}
\begin{tabular}{lcccc} \hline \hline
 & & \hspace{0.3cm} TDLDA \hspace{0.3cm}
 & \hspace{0.3cm} BSE \hspace{0.3cm}
 & \hspace{0.3cm} Exp. \hspace{0.3cm} \\ \hline
TAB & $n\pi^\star$ & 2.1 & 2.5 & 2.80 \cite{lednev98} \\
 \hspace{0.5cm} & $\pi\pi^\star$ & 3.3 & 4.0 & 3.94 \cite{lednev98} \\ \hline
CAB & $n\pi^\star$ & 2.2 & 2.6 & 2.86 \cite{nagele97} \\
 \hspace{0.5cm} & $\pi\pi^\star$ & 3.2 & 3.6 & 4.38 \cite{nagele97} \\ \hline \hline
\end{tabular}
\end{table}

\newpage
\clearpage
\section*{ Figures}

\begin{figure}
\centering\epsfig{figure=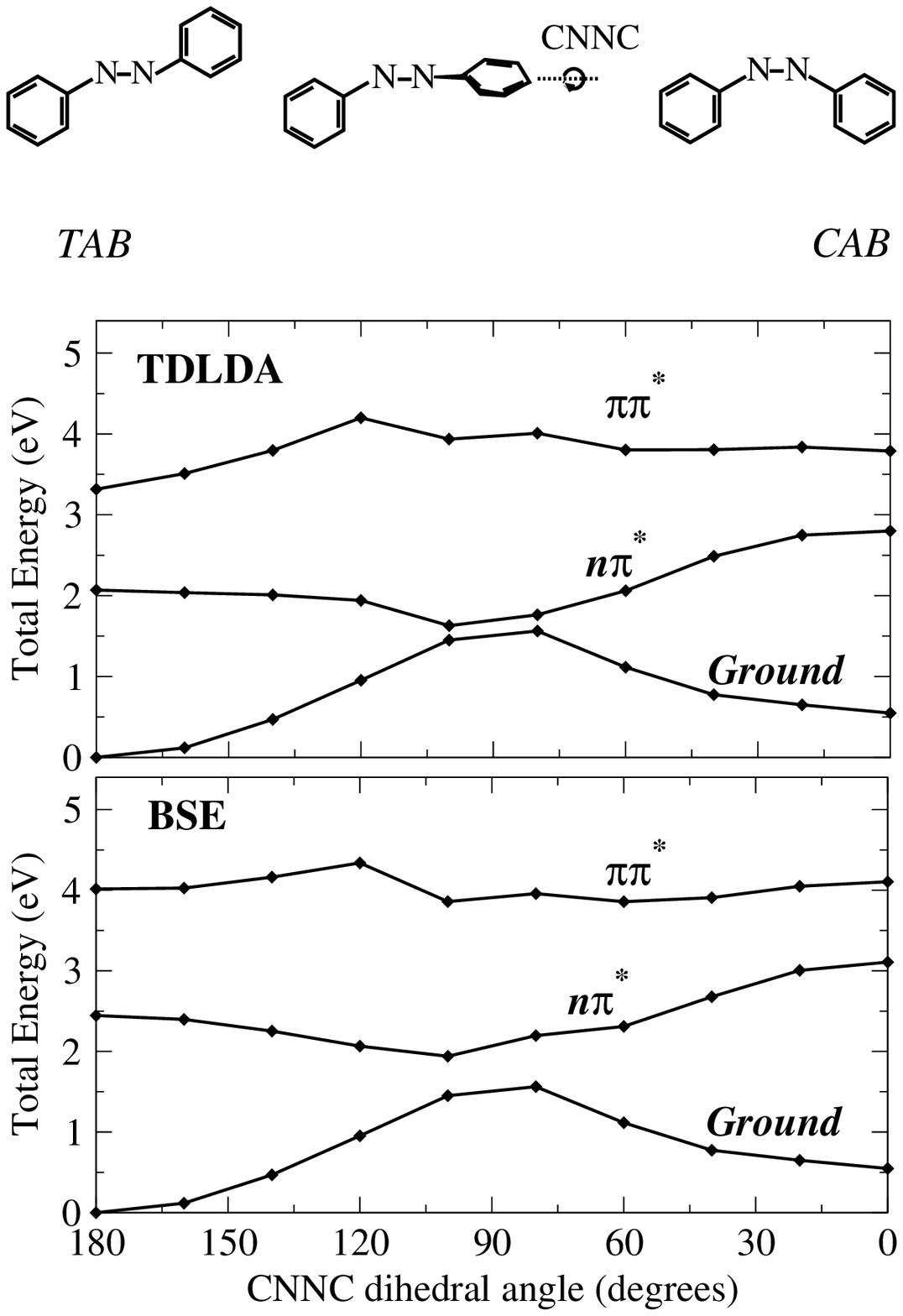,width=12cm,clip=}
\caption{}
\label{f_azo_singlet}
\end{figure}

\begin{description}

\item {Fig. \ref{f_benzene} } Absorption spectrum of benzene,
  calculated within TDLDA (upper panel) and BSE (lower panel). A
  Gaussian broadening of 0.15 eV was used below 10.0
  eV, and 0.5 eV above that frequency. The measured spectrum
  \cite{koch72} is shown in dashed lines.

\item {Fig. \ref{f_azo_singlet} } Potential energy of the isolated
  azobenzene molecule in its ground state and the two lowest
  spin-singlet states, obtained within TDLDA (upper panel), and BSE
  (lower panel). The molecular structure is schematically
  depicted at the two stable configurations (TAB and CAB) and one
  intermediate configuration along the isomerization path.

\end{description}

\clearpage
\newpage

\begin{figure}
\centering\epsfig{figure=benzene.eps,width=12cm,clip=}
\caption{}
\label{f_benzene}
\end{figure}

\clearpage
\newpage

\end{document}